\documentclass[prl,superscriptaddress,groupedaddress,a4paper,twocolumn,nofootinbib]{revtex4}
\usepackage[utf8]{inputenc}
\usepackage[french]{babel}
\usepackage{amsmath,amsthm,amsfonts,amssymb,times,graphicx,color}
\usepackage{multirow}
\usepackage{physics}
\usepackage{braket}
\usepackage{subfigure}
\usepackage{mathtools}
\usepackage{url}
\usepackage{color}
\usepackage{soul}
\mathtoolsset{showonlyrefs}

\include{kkmacros}

\def\GDM#1{{\textcolor{black}{ #1}}}

\def\PA#1{{\textcolor{black}{ #1}}}

\begin{document}

\title{The Grover search as a naturally occurring phenomenon}

\author{Mathieu Roget}
\affiliation{Aix-Marseille Univ, Universit\'e de Toulon, CNRS, LIS, Marseille, France}

\author{St\'ephane Guillet}
\affiliation{Aix-Marseille Univ, Universit\'e de Toulon, CNRS, LIS, Marseille, France}

\author{Pablo Arrighi}
\affiliation{Aix-Marseille Univ, Universit\'e de Toulon, CNRS, LIS, Marseille, France  and IXXI, Lyon, France}

\author{Giuseppe Di Molfetta}
\email{giuseppe.dimolfetta@lis-lab.fr}
\affiliation{Universit\'e Publique, CNRS, LIS, Marseille, France and Quantum Computing Center, Keio University}

\date{\today}
\begin{abstract}
We provide first evidence that under certain conditions, 1/2-spin fermions may naturally behave like a Grover search, looking for \PA{topological} defects in a material. The theoretical framework is that of discrete-time quantum walks (QW), i.e. local unitary matrices that drive the evolution of a single particle on the lattice. Some \PA{QW} are well-known to recover the $(2+1)$--dimensional Dirac equation in continuum limit, i.e. the free propagation of the 1/2-spin fermion. We study two such Dirac QW, one on the square grid and the other on a triangular grid reminiscent of graphene-like materials. The numerical simulations show that the walker localises around the defects in $O(\sqrt{N})$ steps with probability $O(1/\log{N})$\PA{,} in line with previous QW \PA{search on the grid}. The \PA{main} advantage brought by those of this paper is that \PA{they could be implemented} as `naturally occurring' freely propagating particles over a \PA{surface featuring topological---without the need for a specific oracle step. From a quantum computing perspective, however, this hints at novel applications of QW search : instead of using them to look for `good' solutions within the configuration space of a problem, we could use them to look for topological properties of the entire configuration space}.  
%This practical way of performing the oracle step could carry through to experimental implementations of QW on various substrates.
\end{abstract}


%\pacs{}

\maketitle

Quantum Computing has three main fields of applications: quantum cryptography; quantum simulation; and quantum algorithmics (\textit{e.g.} Grover, Shor...). \PA{Some quantum cryptographic devices are already commercialized, and we may hope that some quantum simulation devices will also reach this stage within the next decade. Quantum algorithms, however, are generally considered to be a long-term application.} 
%Whilst the first two are considered short- and mid-term applications respectively, the last one, perhaps the most fascinating, is generally considered to be a long-term application. 
This is because of the common understanding that we will need to build scalable implementations of universal quantum gate sets with fidelity $10^{-3}$ first, and implement quantum error corrections then, in order to finally be able to run our preferred quantum algorithm on the thereby obtained universal quantum computer. This seems feasible, yet long way to go.

In this letter we argue that this may be a pessimistic view. Scientists may get luckier than this and find out that nature actually implements some of these quantum algorithms `spontaneously'. Indeed, the hereby presented research suggests that the Grover search may in fact be a naturally occurring phenomenon, e.g. when fermions propagate in crystalline materials under certain conditions.  

Amongst all quantum algorithms, the reasons to focus on the Grover search \cite{grover_fast_1996} are many. First of all because of its remarkable generality, as it speeds up any brute force $O(N)$ problem into a $O(\sqrt{N})$ problem. Having just this quantum algorithm would already be extremely useful. Second of all, because of its remarkable robustness : the algorithm comes in many variants and has been rephrased in many ways, including in terms of resonance effects \cite{ROMANELLI2006274} and quantum walks \cite{childs2004spatial}.
   
   \begin{figure}[h!]
\includegraphics[scale=0.20]{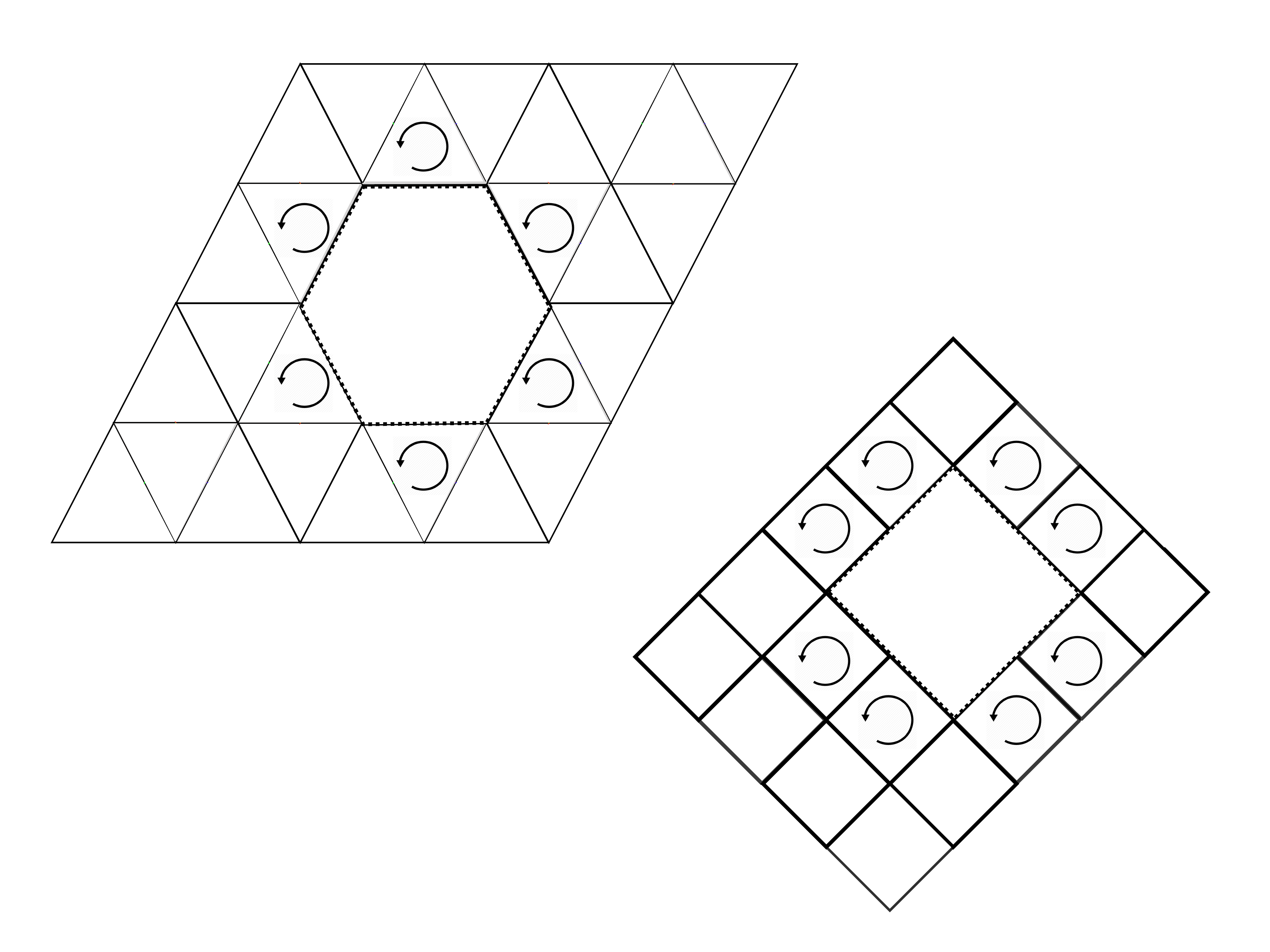}
\caption{Defects in triangular (left) and square (right) lattices. \label{fig:holes}}
\end{figure}
 
   \begin{figure}[h!]
\includegraphics[scale=0.20]{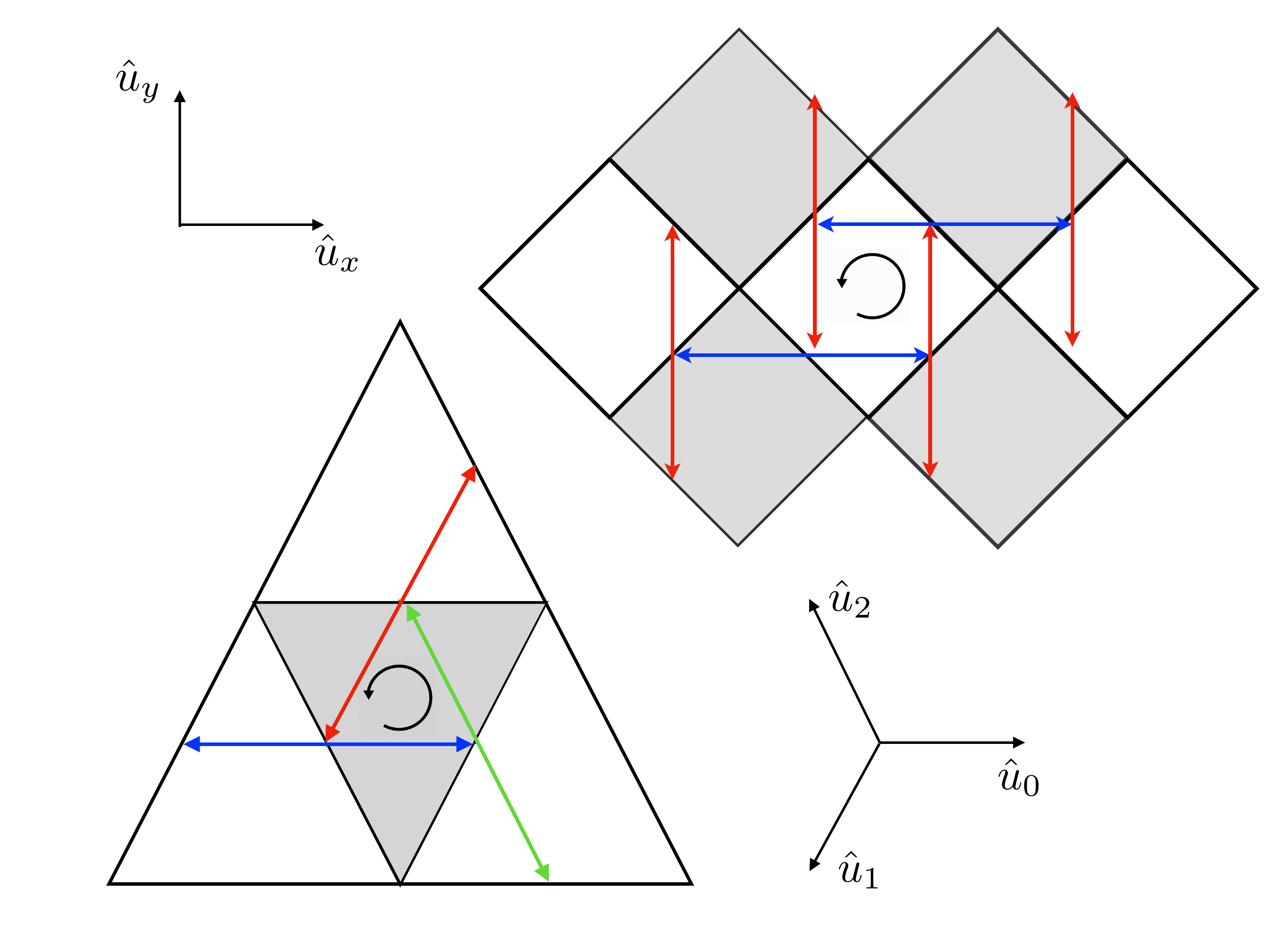}
\caption{Quantum Walks scheme on triangular (left) and square (right) lattice. \GDM{On the triangular grid, an anti-clockwise rotation $R$ is implemented by a partial shift $T_{0,\varepsilon}$ (blue) along a direction ${\bf u}_0$, $T_{1,\varepsilon}$ (red) along a direction ${\bf u}_1$ and $T_{2,\varepsilon}$ (green) along a direction ${\bf u}_2$. On the square grid, an anti-clockwise rotation $R$ is implemented by only two partial shifts $T_{x,\varepsilon}$ (blue) along a direction ${\bf u}_x$, $T_{y,\varepsilon}$ (red) along a direction ${\bf u}_y$. }\label{fig:operator}}
\end{figure}

    Remember that a quantum walks (QW) are essentially local unitary gates that drive the evolution of a particle on a lattice. They have been used as a mathematical framework to express many quantum algorithms e.g.  \cite{BooleanEvalQW,ConductivityQW}, but also many quantum simulation schemes e.g. \cite{di2014quantum, di2016quantum, ArrighiQED}. This is where things get interesting.
      Indeed, it has been shown that many of these QW admit, as their continuum limit, 
    %some well-known PDE of physics, such as 
    the Dirac equation \cite{bialynicki-birula_weyl_1994, meyer_quantum_1996, hatifi2019quantum, di2020quantum},
    %Recall that the Dirac equation governs the propagation of all free spin-$1/2$ fermions. 
  %  Thus, these Dirac QW
   providing `quantum simulation schemes', for the future quantum computers, to simulate all free spin-$1/2$ fermions.
    %, to simulate these particles. 
%    For instance \cite{karafyllidis_quantum_2015} shows that it is possible to describe the dynamics of fermions in graphene using QW. This is great, but now let us turn things the other way round : this also means that fermions provide a natural implementation of these Dirac QW. Could these be useful algorithmically?
%\GDMc{Lately, these Dirac QW have been used to model topological phases effects, such as the fractional quantum hall effect and the presence of edge states $[11,33]$.} %\cite{cardano2017detection, xiao2017observation}.
    
    \PA{Recall that the Grover search is the alternation of a diffusion step, with an oracle step. Here we provide evidence that 1/ these Dirac QW, in $(2+1)$--dimensions, work fine to implement the diffusion step of the Grover search and 2/ Topological defects also work fine to implement the oracle step of the Grover search.}
  %   Thus,  free fermions,
  %  free 1/2-spin fermions \GDM{(or free fermions, or any propagating free quantum particle---as we do not make use of canonical anticommutation relations)} 
   % may provide a natural implementation of this step.
   %  \GDM{Similar conclusions were reached over the square grid \cite{patel2010search}, which we extend to a triangular grid reminiscent of the naturally occurring, graphene-like materials.} 
   %However, recall that the Grover search is the alternation of a diffusion step, with an oracle step. The latter puts on a minus one whenever the walker hits the solution of the problem. Could the oracle step be naturally implemented in terms of 1/2-spin fermions, as well? 
    
    %Here we show 
    %provide evidence 
    %that the mere presence of hole defect suffices to implement an effective oracle step, cf. Fig. \ref{fig:holes}.a. 
    
    \PA{The} second point is, on the practical side, probably more important than the \PA{first}. Indeed, whilst there are several experimental realizations of QW, including 2D QW, 
    %often using cold atoms or photons 
    \cite{WernerElectricQW,Tangeaat3174}, these have not been considered as scalable substrates for implementing the Grover search so far---probably due to the lack of an easy way of implementing the oracle step.
    
    \PA{From a theoretical perspective, that point is strongly suggestive also. Indeed, recall that many quantum algorithms are formulated as a QW search on a graph, whose nodes represent elements the configuration space of a problem, and whose edges represent the existence of a local transformation between two configurations---see \cite{QWSubset} for a recent example of that. So far, the QW search has only been used to look for `marked nodes', i.e. good configurations } within the configuration space, as specified by an oracle. Here, in contrast, the QW search is used to look for topological defects, which are properties of the configuration space itself. This suggests aiming beyond recognizing simple hole defects in 2D crystals, to target more general topological classification problems---e.g. seeking to characterize homotopy equivalence over configuration spaces that represent manifolds as CW-complexes \cite{mermin1979topological, aschenbrenner2015decision}.
%Thus, we have here a first indication that these Dirac QW may be useful to recognize simple hole defects in 2D crystals and eventually be extended to solve more general topological defects classification problems, in terms of mathematical homotopy theory \cite{mermin1979topological, aschenbrenner2015decision}.

%     In a sense this would not be surprising, since it is now a well-established fact that free fermions are subject to topological orders \cite{nomura2007topological, lan2016theory}}.

%    This paper focusses on Dirac QW in $(2+1)$--dimensions, on both the square grid and the triangular grid. 
%%    The triangular grid is of particular interest for instance because of its ressemblance to several naturally occurring crystal-like materials. 
%%    Moreover, it features topological phase effects which, by creating edge states around the hole defect, may help improve localization. 
%    Notice \PA{that }\GDM{non-Dirac }\PA{QW-based }\GDM{Grover search} has already been described on triangular grids in \cite{abal_spatial_2012, chagas_staggered_2018} and more generally on a variety of graphs before. 
%    These spatial search yield $O(\sqrt{N \log(N)})$ time complexity algorithms \cite{aaronson2003quantum,tulsi2008faster, patel2010search}. 
%    The sole aim of this contribution is to show that simple variations of these are in fact naturally occurring \GDM{phenomena}---with the hope to open new and more direct routes towards practical implementations of the Grover search.   

\section{Dirac quantum walks}

%
%\begin{figure}
%{\center
%\includegraphics[scale=0.20]{2D_2.pdf}
%}
%\caption{Quantum Walks scheme on triangular (left) and \GDM{square} (right) lattice. \label{fig:holes}}
%\end{figure}
    
We consider QW both over the square and the triangular grid. More precisely we consider periodic tilings of the plane, where the tiles are either squares or equilateral triangles, of alternating grey and white colours, as in Fig. \ref{fig:operator}. The walker lives over the middle points of each side (aka facet) of each tile. For the square grid we can label these points by their positions in $\mathbb{Z}^2$, for the triangular grid this would be a subset of $\mathbb{Z}^2$. To any such point ${\bf x}$ we assign a complex number representing the amplitude of the walker being there, which we denote by $\psi^+({\bf x})$ (resp. $\psi^-({\bf x})$) if the tile is white (resp. grey). Of course wherever two facets are glued, so are their middle points, and so the two complex numbers form a spinor $\psi({\bf x})=(\psi^+({\bf x})\ \ \psi^-({\bf x}))^\top$ in $\mathcal{H}_2$. Letting $\ket{v_+}=(1\ \ 0)^\top$ and $\ket{v_-}=(0\ \ 1)^\top$ we may then write $\psi({\bf x})=\psi^+({\bf x}) \ket{v_+}+\psi^-({\bf x}) \ket{v_-}$. This degree of freedom at a single point is referred to as the walker's `coin' or `spin'. For the full square grid, the overall state of the walker therefore lies in the composite Hilbert space $\mathcal{H}_2\otimes \mathcal{H}_{\mathbb{Z}^2}$ and can be written $\ket{\psi}=\sum_{{\bf x}} \psi^-({\bf x}) \ket{v_-}\otimes\ket{{\bf x}} + \psi^+({\bf x}) \ket{v_+}\otimes\ket{{\bf x}}$. For the full triangular grid the amplitude of one in every two position needs be zero. For a grid with a missing white (resp. grey) tile the corresponding $\psi^+({\bf x})$ (resp. $\psi^-({\bf x})$) for ${\bf x}$ on a side of the tile, needs be zero.\\
The class of evolution operators that we consider in this paper are QW of the form:
\begin{equation}
\ket{\psi(t+\varepsilon/l)}=WR \ket{\psi(t)}\label{eq:main}
\end{equation}    
with $l=2$ for the square grid and $l=3$ for the triangular grid. Here, $R$ stands for the synchronous anti-clockwise rotation of all tiles. Notice that, wherever there is no missing tile, the simultaneous rotations of the two tiles glued at ${\bf x}$ precisely \GDM{coincides with the implementation of }a partial shift $T_{k,\varepsilon}$ along a direction ${\bf u}_k$:
\begin{equation}
T_{k,\varepsilon} \begin{pmatrix}\psi^+ ({\bf x})\\ \psi^- ({\bf x})\end{pmatrix} =\begin{pmatrix}\psi^+ ({\bf x} + {\bf u}_k \varepsilon)\\ \psi^- ({\bf x} - {\bf u}_k \varepsilon)\end{pmatrix}. 
\end{equation}
Moreover, $W$ stands for the synchronous application of a $2\times 2$ unitary $W({\bf x})$ on the spins $\psi({\bf x})$. This unitary depends on ${\bf x}$ only in a very simple way which we now clarify. First of all, if it so happens that a tile is missing at ${\bf x}$, then the spinor $\psi({\bf x})$ is incomplete, and so $W({\bf x})=I$. Second of all, if there is no missing tile, then $W({{\bf x}})=W_k$, where the $k$ is that corresponding to the partial translation direction ${\bf u}_k$ occurring at ${\bf x}$.\\
It follows that, in the case of a full grid, any given walker will undergo $T_{k,\varepsilon}$ and then $W_k$ for $k=0\ldots l-1$ successively, amounting to
\begin{align}
\ket{\psi(t+\varepsilon)} &=W_{\textrm{$l$}-1}T_{\textrm{$l$}-1,\varepsilon}\ldots W_0T_{0,\varepsilon} \ket{\psi(t)}\nonumber\\
&= \Pi_k W_kT_{k,\varepsilon} \ket{\psi(t)}. \label{eq:main2}
\end{align}
The way we choose these $W_k$ is so that QW is Dirac QW, meaning that
\begin{equation}
\Pi_k W_kT_{k,\varepsilon}\approx \exp(i\varepsilon H_D)  \label{eq:main3}
\end{equation}
as we neglect the second order terms in $\varepsilon$, with $H_D$ the Dirac Hamiltonian in natural $\hbar=c=1$ units, i.e. 
$$H_{D}=p_{x}\sigma_{x}+p_{y}\sigma_{y}+m\sigma_{z}.$$ 
Therefore, on the full grid, these QW simulate the Dirac Equation, more and more closely as $\varepsilon$ goes to zero. 
%, thereby mimicking the free propagation of the 1/2-spin fermion more and more closely as $\varepsilon$ goes to zero. 

\paragraph{Square grid.} Let us consider the unit vectors along the $x$--axis and $y$--axis, namely $\{{\bf u}_x,{\bf u}_y\}$ and use them to specify the directions of the translations $T_{x,\varepsilon}$ and $T_{y,\varepsilon}$. Eq. \eqref{eq:main2} then reads:
 \begin{equation}
U = W_+ T_{y,\varepsilon}W_- T_{x,\varepsilon}
 \label{eq:mainSQ}
 \end{equation}
where $W_\pm = \exp(i \sigma_x \theta_\pm )$ with $\theta_\pm = \pm (\frac{\pi}{4} \pm \varepsilon m)$ and $m$ a is real constant, namely the mass. In the formal limit for $\varepsilon \rightarrow 0$, Eq.\eqref{eq:mainSQ} recovers the Dirac Hamiltonian in $(2+1)$--spacetime. Iterations of the walk observationally converge towards solutions of the Dirac Eq., as was proven in full rigor in \cite{arrighi2014dirac}, \GDM{which motivated the above choice of $U$ on the square grid in the following.}

\paragraph{Triangular grid.} For the Triangular grid let us consider the unit vectors $\{{\bf u}_0,{\bf u}_1,{\bf u}_2\}$, as in Fig. \ref{fig:operator} and defined by
\begin{equation}
        {\bf u}_k = \cos\left({\frac{2 k \pi}{3}}\right) {\bf u}_x + \sin\left({\frac{2 k \pi}{3}}\right) {\bf u}_y  \hspace{0.25cm} \text{for}\hspace{0.25cm} k= 0,1,2.
\end{equation} 
and use them to specify the directions of the translations $T_{i,\varepsilon}$. Eq. \eqref{eq:main2} then reads:
 \begin{equation}\label{eq:mainTR}
        e^{-i\varepsilon H_D} = WT_{2,\varepsilon}WT_{1,\varepsilon}WT_{0,\varepsilon}    
        \end{equation}
with $W = e^{i\frac{\pi}{3}}e^{ -i \frac{\alpha}{2} \sigma_y}e^{- i \frac{\pi}{3}\sigma_z}e^{i \frac{\alpha}{2} \sigma_y}e^{-i\varepsilon\frac{3}{\sqrt{5}}m\sigma_z}$ the coin operator, which turns out not to depend on the direction ${\bf u}_k$. In \cite{arrighi_dirac_2018} it has been proved in detail by some of the authors how this particular choice also leads, in the continuum limit, to the Dirac Hamiltonian in $(2+1)$--spacetime, \GDM{which again motivated us to adopt the above $W$ on the triangular grid in presence of defects.}

\paragraph{Defects.} A sector of a crystal lattice may be inaccessible, e.g. due to surface defects such as the vacancy of an atom (e.g. Schottky point defect) and others. These affect the physical and chemical properties of the material, including electrical resistivity or conductivity \cite{ibach1995solid}. 
%In fact all real solids are impure with about one impurity per $10^6$ atoms. 
Here we model these defects in the simplest possible way: locally, a small number of squares or triangles are missing, thereby breaking the translational invariance of the lattice. In other words, the walker is forbidden access to a ball $\mathcal{B}$ of unit radius, as in Fig. \ref{fig:holes}.This is done by reflecting those signals that reach the boundary $\mathcal{\partial B}$ of the ball, simply by letting $W=I$ on the facets around $\mathcal{\partial B}$. 

Notice that, wherever we replace the coin $W$ by identity, both Dirac walks reduce to just anti-clockwise rotation $R$ as in Eq. \eqref{eq:main}, see Fig. \ref{fig:holes}. Still, the operators \eqref{eq:mainSQ} and \eqref{eq:mainTR} may have different topological properties around $\mathcal{\partial B}$. For instance, the square grid QW has vanishing Chern number and trivial topological properties \cite{kitagawa2012topological} for vanishing $m$, which can still become non-trivial from $m>0$ \cite{asboth2015edge}. The triangular walk on the other hand is always topologically non-trivial and has Chern number equal to one \cite{kitagawa2010exploring}. In the triangular case the positive and negative component decouple respectively in the grey and the white triangles, and may be thought of as inducing polarized local topological currents of spin, called edge states \cite{verga2017edge}. According to \cite{verga2017edge}, this phenomenon will be observed whenever initial states have an overlap with $\mathcal{\partial B}$, elsewhere the walker does not localize and explores the lattice with ballistic speed. Thus, we expect these topological effects to become significant in the triangular case and it is indeed the case. 

Our conjecture is that, starting from a uniformly superposed wavefunction, the walker will, in finite time, localise around the defect in $O(\sqrt{N})$ steps, with probability in $O(1/\log(N))$, with $N$ the total number of squares/triangles. In the following we discuss the numerical evidence we have for such a conjecture. 

\section{Grover search}

\begin{figure}
{\center
\includegraphics[scale=0.35]{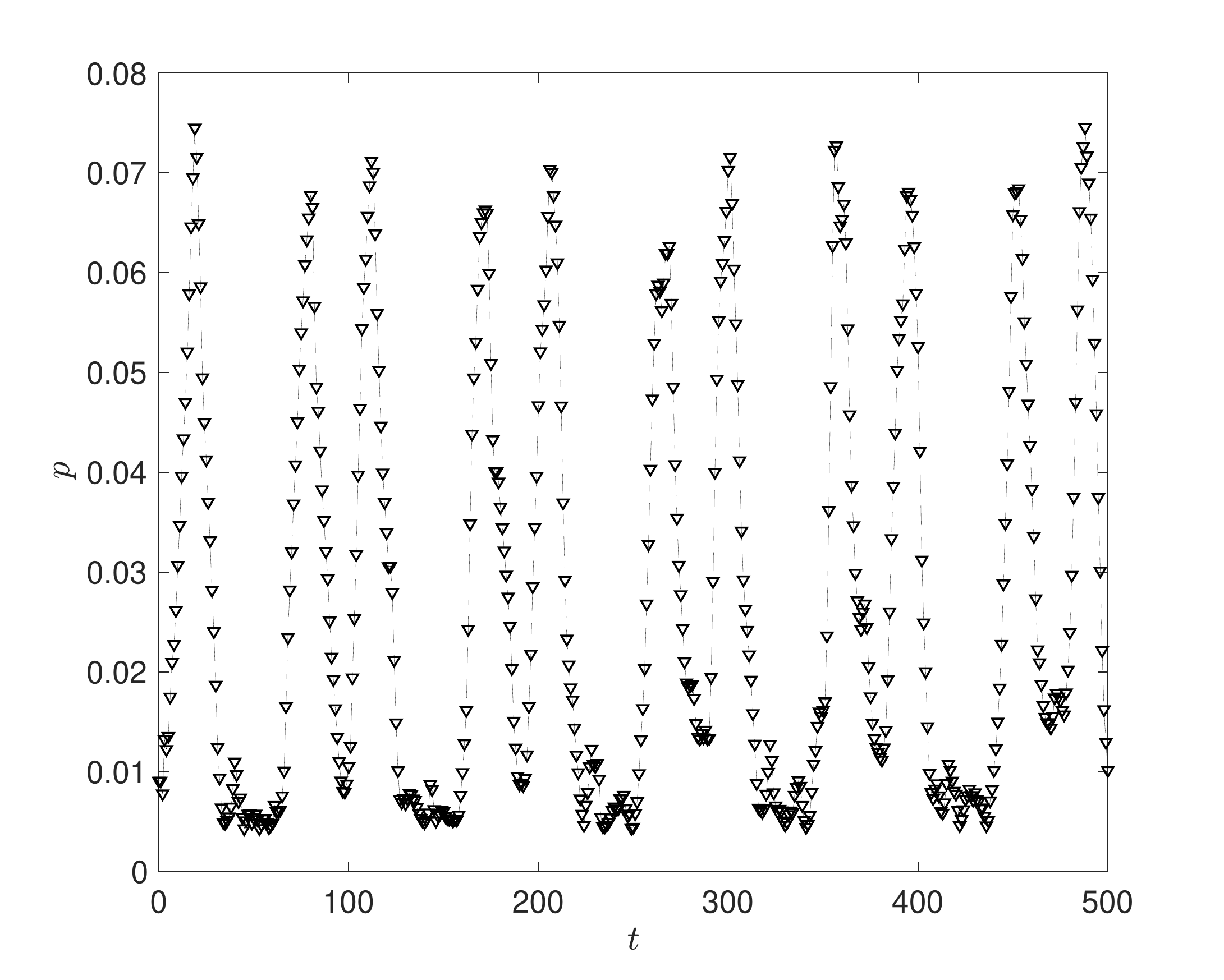}
}
\caption{{\em Square grid periodic localization.} Probability of being localized around the center of the defect versus time. For $m=0$ and $N=2500$.}
\label{fig:perio}
\end{figure}

Our numerical simulations over the square and triangular grids are exactly in line with a series of results \cite{childs2004spatial,aaronson2003quantum,tulsi2008faster} showing that $2D$ spatial search can be performed in $O(\sqrt{N})$ steps with a probability of success in $O(1/\log{N})$. With  $O(\log{N})$ repetitions of the experiment one makes the success probability an $O(1)$, yielding an overall complexity of $O(\sqrt{N}\log{N})$. Making use of quantum amplitude amplification \cite{brassard2002quantum}, however, one just needs $O(\sqrt{\log{N}})$ repetitions of the experiment in order to make the probability an $O(1)$, yielding an overall complexity of $O(\sqrt{N\log{N}})$. This bound is unlikely to be improved, given the strong arguments given by \cite{magniez2012hitting, santha2008quantum,patel2010search}.\\
%. In particular \cite{santha2008quantum,patel2010search} explain why the extra $\sqrt{\log{N}}$, with respect to Grover's original algorithm, is unlikely to be removed.\\
These works were not using Dirac QW, nor defects. Our aim here is demonstrate that QW which recover the Dirac equation, also perform a Grover search, as they propagate over the discrete surface and localise around its defects. More concretely we proceed as follows: (i) Prepare, as the initial state the wavefunction which is uniformly superposed over every square or triangle, and whose coin degree of freedom is also the uniformly superposed $(\ket{v^+}+\ket{v^-})/\sqrt{2}$. Notice that amplitude inside the defect is zero; (ii) Let the walker evolve with time; (iii) Quantify the number of steps \GDM{$t(N)$} before the walker reaches its probability peak \GDM{$p(N)$} of being localized in a ball of radius $2$ around the center of the defect, \GDM{namely the peak recurrence time} and estimate this probability peak, at fixed $N$; (iv) Characterize $t(N)$ and $p(N)$, i.e. the way the peak recurrence time and the probability peak depend upon the total number of squares/triangles $N$.

\begin{figure}
{\center
\includegraphics[scale=0.325]{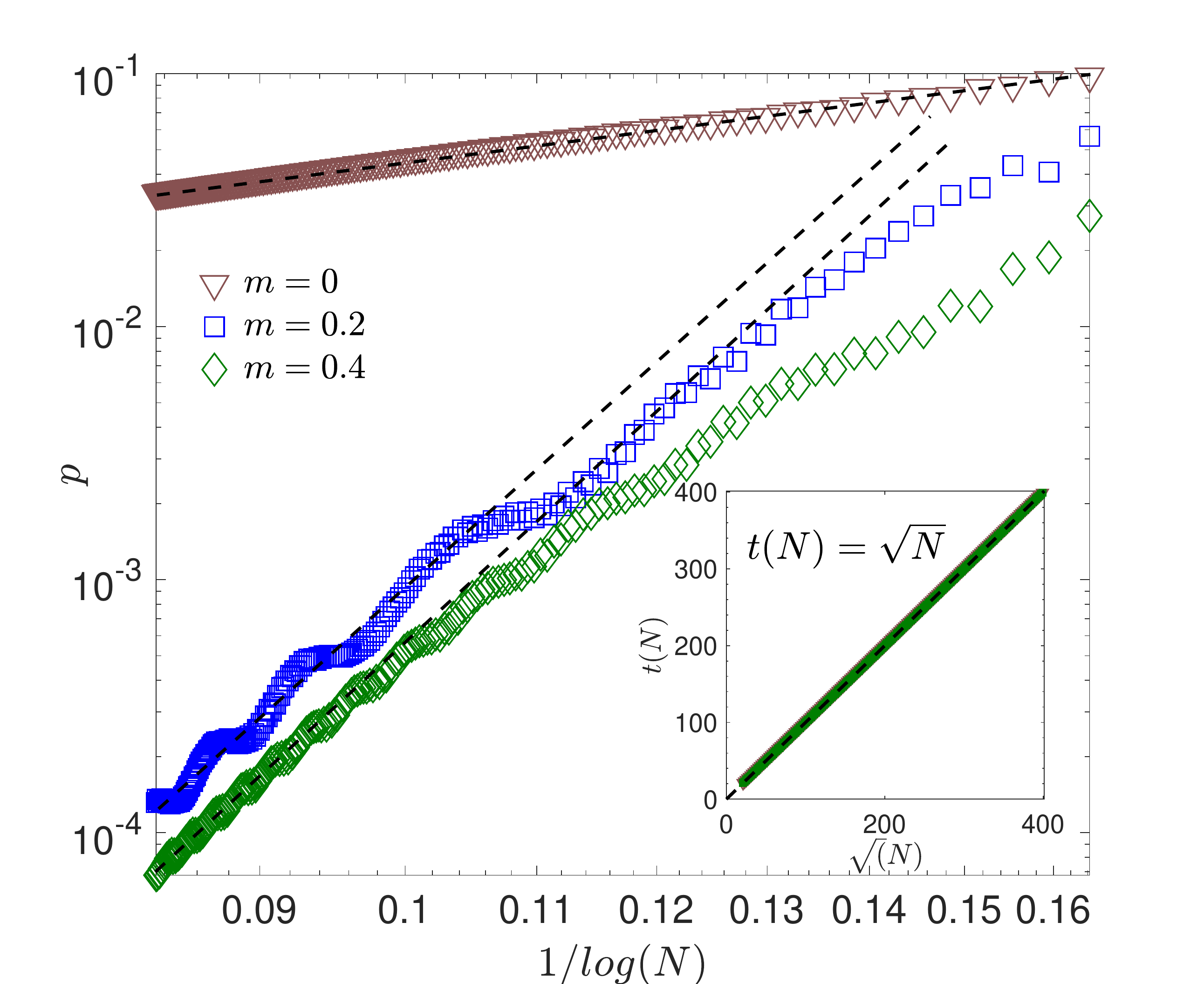}
\includegraphics[scale=0.325]{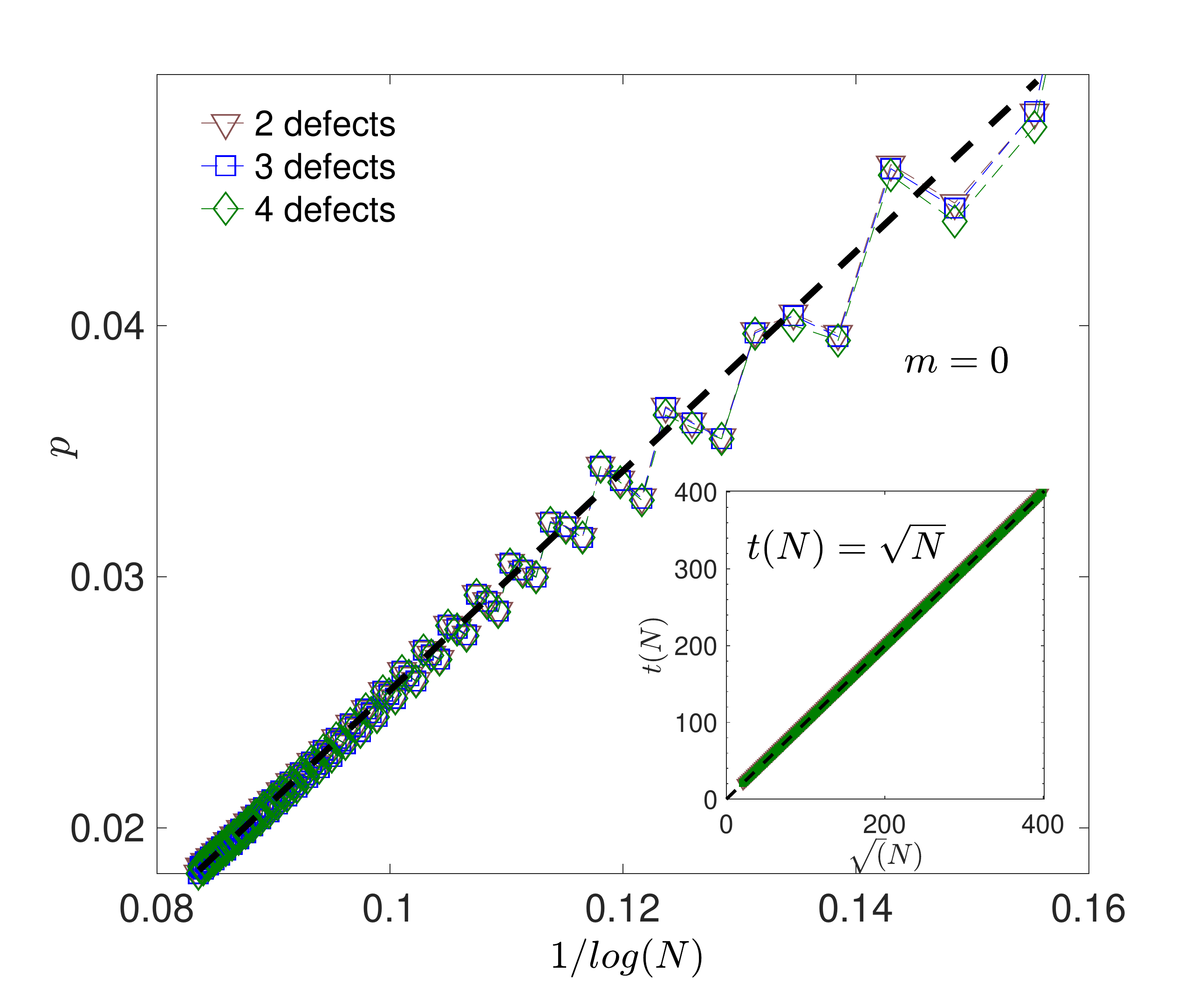}
}
\caption{{\em Square grid scalings}. \GDM{(Top) Probability peak of being localized around one defect, versus the number of squares in the grid for different value of the mass $m$. (Bottom) Probability peak of being localized around two, three and four defects respectively, versus the number of squares in the grid for $m=0$. The inset shows the peak recurrence time.}}
\label{fig:sq}
\end{figure}

We indeed observe that the probability of being found around the defect has a periodic behaviour, see in Fig. \ref{fig:perio} for the case of the square lattice: for instance with $N=2500$ sites, for $m=0$, the peak recurrence time is at $t \sim 25$, with maximum probability is $p \simeq 10^{-1}$. The dependencies in $N$ were interpolated from the data set, as shown in Fig. \ref{fig:sq}.a. We observe that $t(N) = \sqrt{N}$ and $p(N)\simeq 1/\log{N}$ asymptotically, with a prefactor that depends on $m$. In this massless case the interpolation $p(N)\simeq 1/\log{N}$ works right-away, but when the mass gets larger, the curve remains longer along a $p(N)\simeq 1/N$ trajectory, before it eventually enters its asymptotic $p(N)\simeq 1/\log{N}$ regime. \GDM{Moreover, as shown in Fig. \ref{fig:sq}.b, in presence of more than one topological defects we the way the peak recurrence time and the probability depend upon $N$ is the same. Notice that the prefactors do not depend on the number of defects but only depend on $m$, as shown in Fig. \ref{fig:sq}.a.}

Clearly, repeating the experiment an $O(\log N)$ number of times will make the probability of finding the defect as close to $1$ as desired, leading to an overall time complexity in $O(\sqrt{N}\log N)$. Again we could, instead, propose to use quantum amplitude amplification \cite{brassard2002quantum} in order to bring the needed number of repetitions down an $O(\sqrt{\log N})$, leading to an overall time complexity in $O(\sqrt{N\log N})$. But it seems that this would defeat the purpose of this paper to some extent:  since our aim is to show that there is a `natural implementation' of the Grover search, we must not rely on higher-level routines such as quantum amplitude amplification.  

\begin{figure}
{\center
\includegraphics[scale=0.325]{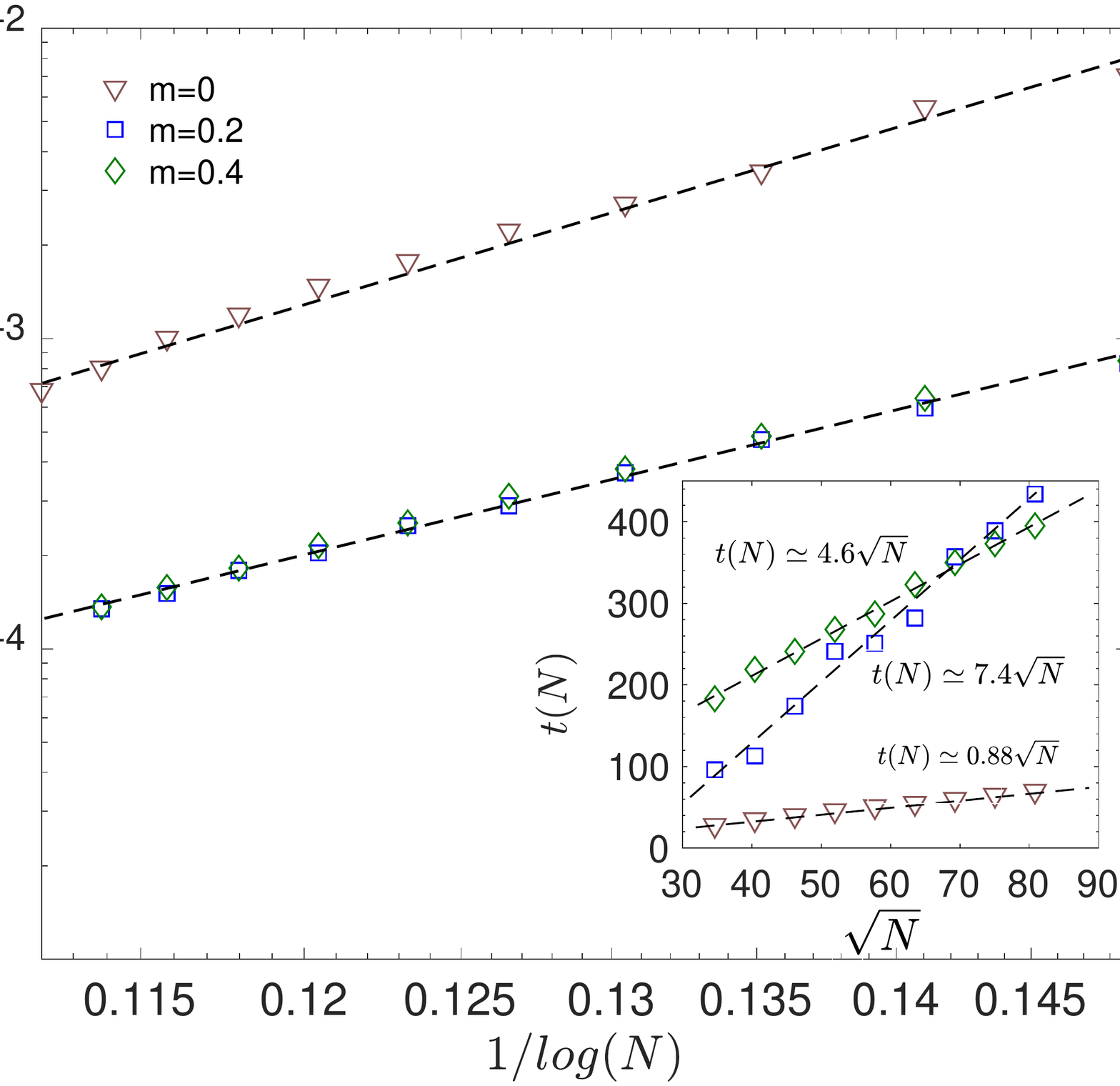}
}
\caption{{\em Triangular grid scalings. } Recurring probability peak of being localized around the defect, versus the number of triangles in the grid. The inset shows the peak recurrence time.}
\label{fig:tr}
\end{figure}

%Over the triangular grid the situation is slightly more intricate, as two phenomena seem to coexist.\\ 
%On the one hand,
Over the triangular grid the Grover search is again at play. Indeed the data set of Fig. \ref{fig:tr}, confirms the results obtained over the square grid: the peak recurrence time is again $t(N)\simeq O(\sqrt{N})$, and its corresponding probability peak is again $p(N)\simeq O(1/\log{N})$ for large $N$, again with a prefactor that depends on the mass. Again this leads to an overall complexity of $O(\sqrt{N}\log{N})$, or $O(\sqrt{N\log{N}})$ using amplitude amplification.\\

\section{Conclusion}

It is now common knowledge that Quantum Walks (QW) implement the Grover search, and that some QW mimic the propagation of the free 1/2-spin fermion.
Yet, could this mean that these particles naturally implement the Grover search? Answering this question positively may be the path to a serious technological leap, whereby experimentalist would bypass the need for a full-fledged scalable and error-correcting Quantum Computer, and take the shortcut of looking for `natural occurrences' of the Grover search instead. So far, however, this idea has remained unexplored. The QW used to implement the Grover diffusion step were unrelated to the Dirac QW used to simulate the 1/2-spin fermion, with the noticeable exception of \cite{patel2010search}. More crucially, the Grover oracle step seemed like a rather artificial, involved controlled-phase, far from something that could occur in nature. This contribution begins to remedy both these objections. 

We used Dirac QW over both the triangular and the square grid as the Grover diffusion step and, instead of alternating this with an extrinsic oracle step, we coded for the solution directly inside the grid, by introducing a \PA{topological} defect. We obtained strong numerical evidence showing that the Dirac QW localize around the defect in $O(\sqrt{N})$ steps with probability $O(1/\log{N})$, just like previous QW search would. 
\GDM{Our next step is to use QW to locate not just a hole defect, but a particular QR code--like defect, amongst many possible others that could be present on the lattice. This would bring us one step closer to a natural implementation of an unstructured database Grover search.}
Replacing the Grover oracle step by surface defects seems way more practical in terms of experimental realizations, whatever the substrate, possibly even in a biological setting \cite{patel2011efficient}. \PA{At a more abstract level, this suggests using QW to search, not just for 'good' configurations within a space, but rather for topological properties of the configuration space itself.}

\paragraph{Acknowledgements} The authors acknowledge inspiring conversations with Fabrice Debbasch, that sparked the idea of Grover searching for surface defects; enlightening discussions on topological effects with Alberto Verga; and useful remarks on how to better present this work by Janos Asboth, Tapabrata Ghosh, Apoorva Patel and the anonymous referees. This work has been funded by the P\'epini\`ere d'Excellence 2018, AMIDEX fondation, project DiTiQuS and the ID\# 60609 grant from the John Templeton Foundation, as part of the ``The Quantum Information Structure of Spacetime (QISS)'' Project. 
%The opinions expressed in this publication are those of the author(s) and do not necessarily reflect the views of the John Templeton Foundation.

\bibliographystyle{plain}	
\bibliography{references}

\end{document}